\pdfoutput=1

\documentclass[conference]{IEEEtran}

\usepackage[spaces,hyphens]{xurl} %
\usepackage[detect-all]{siunitx}
\usepackage[inline]{enumitem}
\usepackage[capitalise]{cleveref}
\usepackage{dirtytalk}
\usepackage{listings}
\usepackage[multiple]{footmisc}

\newcommand{\fbsd}{{F}ree{BSD}}
\newcommand{\krn}{kernel}
\newcommand{\km}{kernel mode}
\newcommand{\mem}{memory}
\newcommand{\oom}{\textsc{OOM}}
\newcommand{\oomk}{\oom\ killer}
\newcommand{\os}{\textsc{OS}}

\newcommand{\um}{user mode}

\crefname{lstlisting}{listing}{listings}
\Crefname{lstlisting}{Listing}{Listings}

\makeatletter
\def\lst@makecaption{
  \def\@captype{table}
  \@makecaption
}
\makeatother

\newenvironment{boxed_new}
    {\begin{center}
    \begin{tabular}{|p{0.45\textwidth}|}
    \hline\\
    }
    {
    \\\\\hline
    \end{tabular}
    \end{center}
    }

\begin{document}

\title{When malloc() Never Returns NULL---\\Reliability as an Illusion}

\makeatletter
\newcommand{\linebreakand}{%
  \end{@IEEEauthorhalign}
  \hfill\mbox{}\par
  \mbox{}\hfill\begin{@IEEEauthorhalign}
}
\makeatother

\author{
  \IEEEauthorblockN{Gunnar Kudrjavets}%
  \IEEEauthorblockA{\textit{University of Groningen}\\
   9712 CP Groningen, Netherlands\\
    g.kudrjavets@rug.nl}
  \and
  \IEEEauthorblockN{Jeff Thomas}
  \IEEEauthorblockA{\textit{Meta Platforms, Inc.} \\
    Menlo Park, CA 94025, USA \\
   jeffdthomas@fb.com}
  \and
  \IEEEauthorblockN{Aditya Kumar}
  \IEEEauthorblockA{\textit{Snap, Inc.}\\
    Santa Monica, CA 90405, USA \\
    adityak@snap.com}
  \linebreakand %
  \IEEEauthorblockN{Nachiappan Nagappan}
  \IEEEauthorblockA{\textit{Meta Platforms, Inc.} \\
    Menlo Park, CA 94025, USA \\
    nnachi@fb.com}
  \and
  \IEEEauthorblockN{Ayushi Rastogi}
  \IEEEauthorblockA{\textit{University of Groningen}\\
    9712 CP Groningen, Netherlands\\
    a.rastogi@rug.nl}
}

\maketitle

\begin{abstract}
For decades, the guidance given to software engineers has been to check the \mem\ allocation results.
This validation step is necessary to avoid crashes.
However, in \um, in modern operating systems (\os), such as Android, FreeBSD, iOS, and macOS, the caller does not have an opportunity to handle the \mem\ allocation failures.
This behavioral trait results from the actions of a system component called an out-of-memory (\oom) killer.
We identify that the only mainstream \os\ that, by default, lets applications detect \mem\ allocation failures is Microsoft Windows.
The false expectation that an application can handle OOM errors can negatively impact its design.
The presence of error-handling code creates an illusion of reliability and is wasteful in terms of lines of code and code size.
We describe the current behavior of a sample of popular \os s during low-\mem\ conditions and provide recommendations for engineering practices going forward.
\end{abstract}


\section{Introduction}

This paper is prompted by observations about how various \os s behave under low \mem\ conditions.
Based on our industry experience, we notice that the application’s \oom\ error-handling code never executes on specific \os s.
We do not observe specific pre-programmed actions such as recovery, retrying allocations, or the presence of relevant log messages.
Instead, applications are just being terminated.
However, recommended software engineering practices encourage engineers to diligently write error-handling code to verify the success of each explicit \mem\ allocation~\cite{mcconnell_code_2004,love_2013,noble_2002}.
Handling the \oom\ errors is supposedly necessary to avoid crashes and ensure that an application continues functioning in a stable state.

We argue that for popular \krn s and \os s based on them, such as Android, Linux, iOS, or \fbsd, the core assumption that an application can reliably handle the \oom\ conditions is outdated.
In most mainstream \os s, \emph{a \um\ application will never have an opportunity to handle a failure to allocate \mem}.
This deviation from a traditional assumption about how to design robust applications is caused by modern \os s using a system component called an \emph{\oomk} (see~\Cref{subsec:oomk}).
Suppose the amount of \mem\ allocated by an application exceeds a certain quota (e.g., a per-process limit) or the amount of free \mem\ for an entire \os\ drops below a certain threshold.
In that case, the \oomk\ will start terminating processes to free \mem.
Processes that will be killed can be an application that fails to allocate \mem\ or any other processes that match the heuristic that the \oomk\ uses.
The presence of an \oomk\ has implications on how reliable applications should be designed and what assumptions they can make about their ability to handle and recover from errors.

We conduct experiments on Android, \fbsd, iOS, Linux, macOS, and Windows to investigate how an application behaves under low-\mem\ conditions.
We also investigate how high \mem\ consumption impacts other applications executing in parallel.
Based on the sample of \os s we investigate; we find that when using default settings:
\begin{enumerate*}[label=(\alph*),before=\unskip{ }, itemjoin={{, }}, itemjoin*={{, and }}]
    \item
    only on Windows, an application can reliably know if \os\ does not have enough \mem\ to satisfy the allocation request and react appropriately
    \item
    applications consuming the most of \mem\ will be terminated before other applications executing in parallel experience the consequences of low-\mem\ conditions
    \item
    though not always a suitable technique, using an \oomk\ is a practical approach to maintain \os 's stability.
\end{enumerate*}
We offer suggestions on how applications should be designed given these limitations and how the presence of an \oomk\ changes the prevailing long-held beliefs about \mem\ management.

\section{Background and motivation}
\label{sec:background}

\subsection{Memory management}

An \os\ is responsible for managing access to various resources exposed to applications.
These resources include \textsc{CPU} time allocation, \mem, storage, and direct hardware access.
Accessing any of these resources is based on an application's \emph{exact} demands.
For example, an application can create a file, allocate a fixed size of \mem, or request the creation of a thread.
Good programming practices for writing reliable code~\cite{mcconnell_code_2004,love_2013,noble_2002} suggest that an application must always check the return value from a mechanism (e.g., a syscall) used to manipulate the resources managed by an \os.
In theory, if allocating resources fails, then an application can decide how to handle this error according to its design principles.
An application can continue to execute in a way that inflicts the least damage to the users or the system's overall stability.

This paper focuses on handling a failure to allocate one category of resources---\emph{\mem}.
The \emph{\krn} manages the entirety of the \mem\ in an \os.
The \krn\ is defined as \say{the part of the system that runs in protected mode and mediates access by all user programs to the underlying hardware (e.g., CPU, keyboard, monitor, disks, network links) and software constructs (e.g., filesystem, network protocols)}~\cite[p.~22]{mckusick_design_2015}.
\emph{User mode} is the counterpart to the software running inside the \krn\ (\emph{\km}).
The scope of the \um\ is limited to \say{programs running outside the kernel}~\cite[p.~298]{tanenbaum_operating_1997}.
\emph{This paper focuses only on \um\ applications}.
Applications such as a browser, a compiler, or a text editors run in \um.
All applications distributed by application stores like Google Play for Android or App Store for iOS and iPadOS execute in \um.
From a typical user's point of view, \um\ applications are mainly what they visibly interact with.

The \krn\ manages \mem\ in chunks of fixed size called \emph{pages} or \emph{page frames}~\cite{tanenbaum_operating_1997,bovet_understanding_2003}.
The \mem\ manager in the \krn\ is responsible for tasks such as accounting for \mem\ usage, managing physical pages, paging, and swapping.
All \um\ applications eventually end up requesting \mem\ from the \krn\ allocator.

One of the \krn 's responsibilities is to ensure the system's overall stability
The \krn\ needs to track the per-process \mem\ allocations as part of this requirement.
If the amount of free physical pages becomes too small, then this threatens the stability of an entire \os.
To alleviate this situation, the \krn\ can terminate \um\ processes to release some of the pages.

The consequences of errors between \km\ and \um\ are different.
The damage caused by incorrect \mem\ management in \um\ is, in most cases, limited to the scope of a single process.
However, a similar mistake in \krn\ mode (e.g., in a device driver) has dire consequences, resulting in a kernel panic~\cite[p.~18]{beck_linux_1998} or a bug check~\cite{russinovich_windows_2012} (more commonly known as {BSoD} or Blue Screen of Death).
As a result, an entire system becomes unusable.

\subsection{The purpose of the \oomk}
\label{subsec:oomk}

The \oomk\ is a built-in facility in some of the \os s responsible for preemptively monitoring the \mem\ usage for an entire \os.\footnote{\protect\url{https://lwn.net/Kernel/Index/#OOM_killer}}\footnote{\protect\url{https://engineering.fb.com/2018/07/19/production-engineering/oomd/}}\footnote{\protect\url{https://source.android.com/devices/tech/perf/lmkd}}
The logic to decide what processes to terminate is based on heuristics such as the type of application, its priority, its \mem\ usage, and a variety of other implementation-specific details.
The \oomk\ uses approaches like monitoring specific watermarks for the number of available and used physical pages of \mem, triggering paging in the \krn, flushing internal caches if needed, and killing processes as a last resort.
The \oomk\ operates within the scope of an individual process.
The necessity to employ an \oomk\ is dictated by the fact that an \os\ cannot assume that each \um\ application manages \mem\ efficiently and does not have \mem\ leaks.
Utilizing the \oomk\ enables an \os\ to continue providing services for all the processes while sacrificing a few.

\subsection{Behavioral differences between \os s}

The general principles of \mem\ management in modern \os s are well-document in literature~\cite{russinovich_windows_2012,anderson_operating_2014,tanenbaum_modern_2001,stallings_operating_2009,singh_mac_2016,love_linux_2005}.
However, different \os s may use conceptually contrasting approaches to \mem\ management that influence the application's design.
For example, we enumerate critical differences between how \krn s in Linux (foundation for Android), iOS, and Windows approach \mem\ management.

\begin{itemize}
\item
Linux uses a concept called \emph{overcommitting}~\cite{love_2013}.
Overcommitting means that when \mem\ is allegedly allocated, a particular \mem\ region is not yet reserved for the application's use.
The actual allocation of \mem\ happens when the allocated pages are modified.
During the write operation, a page access fault is triggered, and only then will \krn\ attempt to allocate \mem.
Linux utilizes the \oomk\ by default.

\item
iOS does not use either \emph{paging} or \emph{swapping}~\cite{levin_ios_2017}.
Paging is an optimization technique to temporarily store the contents of \mem\ on a disk and reload it as needed.
As a result, iOS can only utilize as much \mem\ as physically available.
Memory is one of the most precious resources on iOS.
That design decision forces the iOS \oomk\ to aggressively terminate applications that use too much \mem.

\item
Windows, on the other hand, does not use overcommitting and does not have an \oomk.
The \textsc{NT} \krn\ was designed~\cite{russinovich_windows_2012,richter_2007} to be robust against \um\ applications requesting \say{too much \mem.}
An application can assume that when a pointer to a region of \mem\ is returned, \mem\ will be available.
\end{itemize}

We can see that behavior at the conceptual level between different \os s varies significantly.
These differences, in turn, can have implications on the design assumptions of applications that are intended to run cross-platform.

\subsection{Causes for an \oom\ condition}

An \os\ not having enough \mem\ to complete the desired operation can either be a permanent or a temporary condition.
For example, if a service has a sudden spike in the number of requests it must process, this condition can be transient.
Once the number of requests decreases, the \mem\ consumption will reduce as well.
If during that time, either the service or other applications running in parallel use various mitigation to temporarily deal with \mem\ pressure, then they can continue to execute.

If an application has a consistent \mem\ leak (e.g., a cache that is not bounded or allocations that are never released), then without killing the process, the \os\ will eventually run out of \mem.
If the leak is in a \um\ process, the \mem\ can be eventually reclaimed by terminating the process(es).
If the leak is in the \krn\ itself (e.g., in an \textsc{I/O} manager code), then the only solution to reclaim the \mem\ is to reboot the system.

\subsection{Strategies to handle the \oom\ conditions}

The standard guidance for engineers is to check the result of each call to one of the functions related to \mem\ allocation and assume that they can fail~\cite{mcconnell_code_2004,love_2013,noble_2002}.
This rationale assumes that when \mem\ is unavailable, the allocator will return a value that signifies that the \mem\ request failed.
If \mem\ allocation fails, the engineer can do any of the following:

\subsubsection{Clean up and return}
    The function needs to release the resources already allocated at its scope, roll back the changes in the global state, and return the appropriate error code or an exception to the caller.
    The central assumption here is that the caller will handle the error and perform similar actions.
    Eventually, the user or a top-level caller is communicated the reason why an operation failed.
    The primary intent behind this strategy is to return the system to a stable state and ensure continued execution~\cite[p.~49]{noble_2002}.

\subsubsection{Log an error message and exit the current process}
The basic pattern for this approach is displayed in~\Cref{code:mem_basic_handling}.

\lstset{language=C,breaklines=true,frame=single,basicstyle=\ttfamily\small,caption={Handling an \oom\ condition in {C}.},label=code:mem_basic_handling}
\begin{lstlisting}
void *p = malloc(N);

if (!p) {
    perror("OOM");
    exit(EXIT_FAILURE);
}
\end{lstlisting}

    The rationale for this behavior is that if no more \mem\ can be allocated, further use of the application is undesirable.
    The best course of action is to record what happened and hope restarting the application will avoid the problem next time.
    The application can also try to flush the pending changes to the disk, save the current state, or perform  other actions to help with future recovery.
    There are no guarantees, however, that any actions, including logging the error message itself (that may require allocating \mem), will succeed.

    \subsubsection{Release and retry}
    The application can perform a last-minute attempt to free any resources available (e.g., release \mem\ allocated for an internal cache) and then retry allocating \mem.
    The reasoning is that if the application stores many objects in the \mem, then the cache can be populated later, and the freed \mem\ may help the application continue executing.

    \subsubsection{See no evil, hear no evil, speak no evil}
    Approaches like \emph{ignoring the problems related to resource management}~\cite[p.~223]{noble_2002}, most famously applied to earlier versions of \textsc{UNIX}~\cite{unix_design}, are also possible.
    Some of the anecdotal reasoning we have heard in the industry is based on the cost of writing reliable code versus the probability and consequences of a crash in non-critical \um\ applications.

\subsection{Strategies to anticipate the \oom\ conditions}

We have observed various techniques that applications use to avoid getting into the \oom\ situation in the first place.
One of the strategies involves \emph{periodically polling the available \os\ \mem} (either a percentage or a fixed size).
Based on the amount of available \mem, an application will preemptively take various steps to reduce the possibility of the \oom.
However, modern \os s utilize concepts such as \emph{cgroups} in Linux~\cite{jain_linux_2020} and \emph{job} objects in Windows~\cite{russinovich_windows_2012,richter_2007}.
Those constructs allow controlling one or more processes as a group.
For example, a creator of a job object can specify how much \mem\ a single process can allocate or how much \textsc{I/O} the process can perform.
In that state, an application can no longer use data about the global state to make correct decisions about \mem\ availability.

\emph{Subscribing to low-\mem\ notifications} is another way to be notified about \mem\ pressure.
Most modern \os s enable applications to react to a situation when the number of free pages falls under a certain threshold.
An application can then try to release resources allocated by it and hope that the resulting impact on the amount of available \mem\ is significant enough for the \oomk\ not to terminate the process.
However, there are no guarantees that any effort performed by an application at this stage will be sufficient to avoid termination.

\emph{Anticipatory Memory Allocation}~\cite{sundararaman_2012} is a technique used to make \krn\ robust to \mem-allocation failures.
Similar logic can be applied to the \um\ applications.
Applications can preemptively estimate how much \mem\ they need and attempt to pre-allocate this amount during the startup.
For non-critical applications, this approach is very wasteful because most of the \mem\ will be hoarded and not used.
Similar methods use \mem\ allocation rate as a predictor~\cite{nakagawa_2015}.

Attempt to \emph{outsmart the \oomk} is another possibility.
The source code for \krn s of \os s such as various distributions of Linux or \fbsd\ is public.
The inner workings of the \oomk\ can be inferred from that source code.
Even though iOS is a closed-source \os, the internals of how Darwin and Mach kernels are implemented are also documented to some degree~\cite{singh_mac_2016,levin_ios_2017}.
\emph{An application can reverse engineer the algorithm the \oomk\ uses to avoid being terminated}.
However, each \os\ can change both the design and implementation of the \oomk\ with each new release or an update, rendering all preemptive measures an application has put in place obsolete.

\section{Empirical findings}

\Cref{code:mem-eater} contains the essential part of a program\footnote{\url{https://figshare.com/s/8ed507efe7d4ed0f6480}} we use on different \os s to gather empirical data about what happens when an application tries to continue allocating \mem\ without releasing it.
The algorithm
allocates a fixed size of \mem,
dirties the allocated pages by explicitly writing into them post-allocation, and
returns an error when \mem\ no longer can be allocated.
The extra step of dirtying the pages is necessary to prevent the optimizations related to overcommitting
and, in some cases, compilers optimizing away the allocation code.

\lstset{language=C,breaklines=true,frame=single,basicstyle=\ttfamily\small,caption={Algorithm to continously consume \mem.},label=code:mem-eater}
\begin{lstlisting}
/* Any size can be used here. */
unsigned alloc_size = 1024 * 1024 * 10;

while (1) {
    void *p = malloc(alloc_size);

    if (!p) {
        perror("OOM!");
        exit(EXIT_FAILURE);
    }

    /* Dirty the allocated pages. */
    memset(p, 'X', alloc_size);
}
\end{lstlisting}

We execute this application on Android, \fbsd, iOS, Linux, macOS, and Windows.
Both as a single instance and multiple copies in parallel with varying allocation sizes.
We observe distinct types of behaviors:
\begin{enumerate*}[label=(\alph*),before=\unskip{ }, itemjoin={{, }}, itemjoin*={{, and }}]
    \item
    \oomk\ terminates only the offending application because it consumes most of the \mem
    \item
    \oomk\ terminates a set of processes (not necessarily including the offender) based on its heuristic (e.g., background versus foreground, priority, recent usage)
    \item
    global alerts from an \os\ indicating lack of \mem\ that result in a user being presented with a choice of what process to manually terminate
    \item
    \textit{malloc()} returns \texttt{NULL} when an \oom\ condition happens.
\end{enumerate*}

This behavior matches what we expect based on inspecting the source code for various implementations of an \oomk\ and existing documentation.
Samples of the output from various \os s are displayed in~\Cref{shell:oom-process-killed}.

\lstset{language=C,breaklines=true,frame=single,basicstyle=\ttfamily\small,caption={Samples of behavior on various \os s.},label=shell:oom-process-killed}
\begin{lstlisting}
# Microsoft Windows 10.0.22000.675
C:\Temp>oom.exe
OOM!: Not enough space

# Ubuntu 20.04.3 LTS (Focal Fossa)
$ ./oom
Killed

# macOS Monterey Version 12.3.1
$ ./oom
zsh: killed     ./oom
\end{lstlisting}

Our experiments are run with default settings.
We do not modify any parameters specific to overcommitting, process-related quotas, or use custom \um\ \mem\ managers.
Default \krn s are used for all the \os s.

\begin{boxed_new}
Out of all the \os s we experiment with, in case of an \oom\ condition, \textit{malloc()} returns \texttt{NULL} only on Windows.
On all the other \os s, a set of processes consuming most of the \mem\ (or meeting the criteria a particular \oomk\ uses) are terminated by the \oomk\ first.
\end{boxed_new}

\section{Discussion and implications}

\subsection{Implications of current state in \mem\ management}

The problem of \say{malloc() never returns NULL,} and shortcomings of the \oomk\ have been known since it was introduced to Linux~\cite{jang_2008,patare_2015}.
Current historical assumptions about handling the \oom\ conditions have a variety of adverse side-effects when it comes to application reliability.

\subsubsection{Lack of ability to handle errors}
When writing code, engineers assume the possibility of recovering and executing of error-handling code.
For most of the popular \os s, this assumption is incorrect.
In low-\mem\ conditions, an application may be killed by the \oomk\ using \texttt{SIGKILL}.
\texttt{SIGKILL} (colloquially known as \texttt{kill -9}) is a signal that causes an application's immediate termination.
Because the application cannot handle, intercept, or block \texttt{SIGKILL}, \emph{an application must assume that it can be terminated at any moment, with or without a cause}.
An application may never have a chance to handle an \oom\ condition in its code.
As a result of sudden termination, an application may leave behind various residues.
For example,
an application could fail
not to clean up the temporary cache of files on a disk,
not release cross-process synchronization primitives,
or even corrupt data.
Therefore, if an application is designed under the assumption that it will have a chance to handle an \oom\ condition, then it needs to be redesigned to match the current reality.

\subsubsection{Lack of control over the execution environment}
In general, most applications do not fully control their execution environment.
That is especially true for mobile applications on Android or iOS, where \os\ \say{sandboxes} the application.
The \os\ will prevent an application from making changes to the device and even querying the data about other applications.
Potential workarounds to avoid being terminated, such as modifying the \oomk\ settings, are not possible by design.

\subsubsection{Lack of fairness}
Ensuring a system's stability is not an entirely fair process.
An application must not be the most significant \mem\ consumer to be killed.
Another process of a higher priority may be using more \mem, and the application may end up on the kill-list regardless.
Alternatively, it may be one of the many applications that are terminated to ensure that the \os\ can continue providing services.
Even if an application releases all the \mem\ possible as a response to a low-\mem\ notification and behaves as a \say{good citizen} in a specific ecosystem, it can still be terminated.

\subsubsection{Maintenance costs}
An application's code base will contain error-handling code that is never executed.
That extra code needs to be maintained and tested.
The presence of extra error detection and recovery code decreases the clarity of the code base and its readability~\cite{holzmann_2015}.
The error-handling code is shown to be a \say{substantial source of faults in systems code}~\cite{saha_2013}.
Other negative side-effects of extra code include increases in the application's size.
Application size, in turn, is one of the factors that need to be tightly controlled on mobile \os s such as Android or iOS~\cite{hort_2021}.
From a testing point of view, additional execution paths will require developing more test cases and effort to reach higher code coverage.

\subsection{Discussion points}

\subsubsection{Overcommitting}

For server \os s, such as different distributions of Linux and BSD, there are several options to control an \os 's behavior.
An organization that deploys applications internally can either build a custom \krn\ or control different settings related to overcommitting.
For example, the Linux \krn\ can be configured to disable overcommitting.
One of the risk factors with this approach is that even if a particular application is designed to handle the \oom\ events properly, the behavior of all the other applications is unknown.

\subsubsection{Multiple platforms}

If an application's code is intended to run cross-platform on \os s without the \oomk, it must handle the possible failures to allocate \mem.
Ideally, a version compiled for each platform will be optimal, i.e., no unnecessary checks for allocation failures if an \oomk\ is present on a target \os.
However, there is a downside to initially omitting the error-handling code.
An application that is originally designed to run only on one \os\ and later ported to a different \os\ will require revisiting all the instances in the code base where \mem\ allocation is performed.

\subsubsection{Exclusion from \oomk}

There is a possibility that applications can exclude themselves from being killed by the \oomk.
However, that option is not available for applications installed through an official mobile application deployment platform such as Apple Store or Google Play Store.
For example, in Linux, an ability to exclude a specific process from being killed may be available only for privileged processes depending on the \krn\ version.
Enabling this approach also introduces the \say{what if everyone did that} dimension to the \oom\ management issue, rendering the \oomk\ ineffective.

\subsubsection{Silent \oomk\ makes debugging time-consuming}

Our observations from the industry indicate that users have a poor experience when the \oomk\ does its job.
We had to debug a multitude of issues when under low-\mem\ conditions, a process experienced a sudden termination.
The actions of the \oomk\ result in a sudden \say{application death} caused by \texttt{SIGKILL}.
However, for the user, there is no clear indication of why the application was terminated.
This problem is well-known in the Linux community.\footnote{\protect\url{https://lwn.net/Articles/894546/}}
Anecdotal evidence from users after enabling \textit{systemd-oomd} by default on Ubuntu Desktop supports this observation.\footnote{\protect\url{https://lists.ubuntu.com/archives/ubuntu-devel/2022-June/042116.html}}

\subsection{Recommendations}

We recommend that engineers adopt the following guiding principles when designing their applications:

\begin{enumerate}

\item
\emph{On \os s that utilize an \oomk, design applications with an assumption that they can be terminated at any time without having a chance to react}.
This concept is not new.
The desire for an application to gracefully recover after a crash has been advocated for in the past~\cite{candea_2003}.
Some applications use patterns such as a global exception handler for unhandled exceptions.
The code that is part of a global exception handler can perform any necessary actions before termination.
However, global exception handlers will not execute when the \oomk\ terminates a process.

\item
\emph{On \os s where an application can subscribe to low-\mem\ notification events, an application should assume that its termination is imminent if notified}.
The best-case scenario in such situations is to treat this notification as an opportunity to either trigger the controlled shutdown sequence or execute a best effort to avoid conditions that would prevent the application's restart.
An application should try to commit pending changes, synchronize its in-\mem\ data structures to the disk, and possibly, as a last-effort attempt, try to release as much \mem\ as possible.

\item
\emph{If an application is intended to be deployed on \os s where the application can handle an \oom\ condition, then it should use a consistent design pattern such as \say{clean up and return} everywhere}.
Utilizing a consistent design pattern where each function is responsible for error-handling and cleanup is a design strategy that requires engineers to exercise this practice meticulously.
However, unless all the dependencies follow a similar approach, this solves only a part of the problem.
Given the engineering cost, we recommend this approach for critical \um\ applications such as daemons or services, depending on the nomenclature a particular \os\ uses.

\item
\emph{If the design of an application supports recovery in case of sudden termination, then we recommend that for \mem\ management, the application switch over to functions such as \textit{xmalloc()},\footnote{\url{https://www.freebsd.org/cgi/man.cgi?query=xmalloc}} \textit{xrealloc()}, and \textit{xfree()} or their equivalents}.
These functions are guaranteed to either return successfully or terminate the application with an appropriate error message.
Using the \say{x-family \mem\ allocation functions} enables the application to omit the error-handling code and simplify the error-handling strategy~\cite{holzmann_2015}.

\item
\emph{If an application is developed in {C++}, the standard library enables interception of a situation when the \texttt{new} operator cannot allocate \mem}.
If \textit{malloc()} will return \texttt{NULL},
an application can choose how to act under the \oom\ conditions in a central location by setting a custom \emph{new-handler}.\footnote{\protect\url{https://en.cppreference.com/w/cpp/memory/new/set_new_handler}}
The handler can try to release some \mem\ and retry the allocation, terminate, or perform a different action depending on the application's design.

\end{enumerate}

\section{Conclusions and future work}
\label{sec:conclusions}

Universally checking the result of a request to allocate \mem\ has been a standard practice for decades.
Our recommendation to ignore that guidance on a subset of \os s is clearly contrarian.
However, software development practices need to adapt to a new reality.
That new reality means, for example, in the case of popular mobile \os s such as Android and iOS, an application is not in control of what happens in case of an \oom\ event.
The typical desktop applications that execute in non-administrative mode have the same limitations.
They cannot change the \os\ settings, query the details about the \mem\ usage of other applications, and cannot circumvent an official \oomk\ to prolong their existence.
\emph{As a result, all the code that is supposed to execute when an \oom\ condition happens will never run.
Therefore, there is no reason for that code to be present}.

One topic for the future work we intend to pursue is the effectiveness of low-\mem\ notifications on \os s that enable them.
We want to study
\begin{enumerate*}[label=(\alph*),before=\unskip{ }, itemjoin={{, }}, itemjoin*={{, and }}]
    \item how many and what types of applications use those mechanisms
    \item what actions do they perform (e.g., is the \mem\ being released or an event is just logged)
    \item how efficient those actions are (e.g., what percentage of \mem\ that an application is responsible for is released)
    \item what is the impact of those actions (e.g., in what percentage of cases the \oomk\ will let an application continue to execute).
\end{enumerate*}

Another subject we are interested in studying is engineers' belief systems about \mem\ management.
Based on our observations, the beliefs depend on the abstraction level the engineers work at.
Engineers working lower in the stack (e.g., compilers, \krn, systems software in general) tend to be more cognizant of the consequences of \mem\ allocation failures.
They believe in doing everything possible to prolong the application's lifetime.
We are interested in studying if there is validity to our observations across the industry.

\section{Threats to validity}
\label{sec:threats}

Like any other study, the results we present in this paper are subject to specific categories of threats~\cite[p.~222--223]{shull_guide_2008}.

Threats to \emph{external validity} are related to application of our findings in a different context.
There are a variety of \os s in existence.
Our experiments were run only on a subset of \os s, albeit the most popular ones.
The behavior of various \krn s and \oomk s is constantly evolving.
Our conclusions are drawn only from a sample of available data.
For commercial \os s, we rely on publicly accessible information~\cite{russinovich_windows_2012,singh_mac_2016,levin_ios_2017,richter_2007} about how their \krn\ behaves.
However, that logic can change during any subsequent releases.

One threat to \emph{conclusion validity} is related to the fact that our reasoning is drawn mainly from our experiences with the development of commercial system software.
That implies a bias towards optimizing certain characteristics of software (e.g., performance cost, presence of unnecessary lines of code) and making trade-offs differently than, for example, in the case of open-source software.

\bibliographystyle{IEEEtran}
\bibliography{oom}

\end{document}